\documentclass[twocolumn,showpacs,amsmath,amssymb]{revtex4}
\usepackage{graphicx}
\usepackage{amsfonts}

\begin{document}

\title{A two dimensional model for ferromagnetic martensites}

\author{S. Sreekala$^1$, and G. Ananthakrishna$^{(1,2)}$
\footnote{Electronic Mail: garani@mrc.iisc.ernet.in}
}
\affiliation{$^1$ Materials Research Centre, Indian Institute of Science,
Bangalore-560012, India\\
$^2$Centre for Condensed Matter Theory, Indian Institute of Science, Bangalore-
560012, India\\
}

\date{\today}

\begin{abstract}
We consider a recently introduced 2-D square-to-rectangle  martensite model 
 that explains several unusual features of martensites to study ferromagnetic
 martensites. 
The strain order parameter is coupled to the magnetic order parameter through
 a 4-state clock model.  Studies are carried out for several combinations of
 the ordering of the Curie temperatures of the austenite and martensite phases 
and, the  martensite transformation  temperature.  We find that the 
orientation of the magnetic order  which  generally points along  the short 
axis of the rectangular variant, changes  as one crosses the twin or the
 martensite-austenite interface.   The model shows the possibility of a 
subtle interplay between the growth of strain and magnetic order parameters as
 the temperature is decreased. In some cases, this  leads to qualitatively
 different magnetization curves from those predicted by earlier mean field 
models. Further, we find that strain morphology can be substantially altered by
 the magnetic order. We have also  studied the dynamic hysteresis behavior.
 The corresponding dissipation  during the forward and reverse cycles has 
features similar to the Barkhausen's noise.
\end{abstract}


\maketitle

\section{Introduction}
 
Mixed order parameter (OP) systems often exhibit unanticipated material 
properties that can be of technological  importance.  A good example is that 
of the recently studied ferromagnetic martensite such as the Heusler alloy.
  Apart from the martensite transformation temperature, such  alloys can in
 principle have different  Curie temperatures for the parent and the product
 phases that  could be higher or lower than the martensite transformation
 temperature. It is believed that an interplay between the associated 
magnetic, elastic and magnetoelastic contributions  leads to  the unique 
 properties of these alloys.   For example, Heusler alloys show reversible 
deformations  that can be induced by application of magnetic field 
 \cite{James02,yamamoto}. Ullakko et al \cite{Ullakko96} were the first to
 report the field induced strain in Heusler alloy and suggest that the  
magnetic field induced strain observed in martensite phases arises from the 
twin boundary motion rather than magnetostriction \cite{likhachev}. Similar
 effect has been reported in $Fe_7Pd_3$ also \cite{james98}. Later 
much  higher  strains  $\sim 6 - 7 \% $ have been reported in Huesler alloys 
\cite{Ullakko96,ohandley98,ohandley00,murray00,harshdeep}.    For this reason,
 ferromagnetic  martensites are  good candidates  for actuator applications.

A large volume of experimental work has been devoted  to study various physical
 properties characterizing the magnetoelastic transition 
\cite{steuwer,wada,murakami,oliver,heczko,cui,chernenko98} including their
 dependence on  alloying composition as the latter affects the transition 
temperatures \cite{vasilev,chernenko}. In view of their technological
 importance,  several models have been proposed to explain  the 
large reversible strain \cite{ohandley98,james98,ohandley00,paul03}. 
Apart from this, there have been other studies to obtain the magnetic 
phase diagrams for different sequences of ordering  of the martensite
 transformation temperature  and the Curie temperatures of the austenite and
 martensite phases  using mean field models 
\cite{buchel1,buchel2,bucheljmmm,chernenko,vasilevcond,vasilev}. 

During martensitic  transformation, the higher symmetry phase undergoes a 
spontaneous deformation  to a lower symmetry phase, for instance from a 
cubic parent phase to  a tetragonal martensitic phase as in the case of
 $Ni_2MnGa$ and $Fe_7Pd_3$. Thus, the short  crystallographic axis is always 
the easy direction of magnetization \cite{straka} and magnetic anisotropy
 can be quite large in some cases. For example, in Ni-Mn-Ga martensites, 
the  magnetic anisotropy in the martensitic phase is ($10^3  J/m^3$ ) 
two orders  greater than the parent austenite phase \cite{tickle}. Thus, 
the spin ordering is expected to change across the twin variants of the
 martensite. However, there has been few attempts \cite{james98} to study 
the changes in  orientation of the magnetic order across the variants of 
the martensites, and also across the  martensite - austenite interfaces.

The purpose of the present paper is  two fold. First, to propose a model that
 deals with the martensite transformation and the magnetic transition on a 
similar footing with the aim of  capturing the influence of magnetic order
 on strain and vice versa.  Second, to study the orientation of the magnetic
 ordering   in the martensite variants and in the parent phase for 
various possible orderings of  the Curie temperatures relative to the 
martensite start temperature. This exercise will be carried out within the
 context of a recently introduced model for square-to-rectangle martensite 
transformation. This model has been shown to capture several unusual
 properties associated with athermal martensites such as the power law
 statistics of the acoustic emission (AE) signals, the correlated nature of
 AE signals when the system is cycled in a small range of temperatures and the
 associated shape memory effect, and the precursors effect 
\cite{vives,muto,robertson,kartha,oshima,rajeev,kalaprl,kalaprb}. 
The model is coupled to  the magnetic order parameter through a four state 
clock model. The clock model is used  as  continuous spin system in two 
dimension do not support long range order. While the strain order parameter is 
governed by a Bales and Gooding type of equation \cite{bales}, the development 
of magnetic order is carried out using Monte-Carlo method.   We find that the 
magnetic order parameter which is generally oriented along  the short axis of
 the rectangular variant changes  as one crosses the martensitic twin 
interface or martensite-austenite interface. We also find that the magnetic
 order influences the strain and vice versa substantially. We also study the 
'noise' associated  magnetic hysteresis \cite{planeshyst,madanhyst,garcia}
 that is similar to Barkhausen noise.

\section{Structural and Magnetic Transformations}
Let $T_{Ms}$ and  $T_{Mf}$ be the martensite start and  finish temperatures
 respectively. Defining  ${T_c}^A$ and ${T_c}^M$ to be  the Curie temperatures 
of austenite and martensite  phases, we have  six different  possible ways of 
ordering of the above temperatures. They are\\
a. ${T_c}^M \le T_{Ms} \le {T_c}^A $ \\
b. $T_{Ms} \le {T_c}^M \le {T_c}^A $ \\
c. $T_{Ms} \le {T_c}^A \le {T_c}^M $ \\
d. ${T_c}^M \le {T_c}^A \le T_{Ms} $ \\
e. ${T_c}^A \le T_{Ms} \le {T_c}^M $ \\
f. ${T_c}^A \le {T_c}^M \le T_{Ms} $ \\
The first four cases are situations where the austenite phase is the first 
to order ferromagnetically.  For the cases $e$ and $f$, the onset of the 
ferromagnetic temperature is determined by $T_{Ms}$ and ${T_c}^M$ respectively.
 Condition $a$ is more interesting because on lowering the temperature, the 
para to ferromagnetic 
transition  takes place in the austenite phase followed by a martensitic 
transformation which  becomes magnetic  only on further lowering the 
temperature.

\section{The Model}

The total free energy of a  ferromagnetic martensite has contributions 
arising  from the elastic energy $F_e (\epsilon)$, magnetic energy 
$F_m(\theta )$, and  magnetoelastic coupling $F_{me}(\epsilon,\theta)$ 
 where  $\epsilon$ is the  order parameter for ferro-elastic transformation and
 $\theta$  the order parameter for the para to ferromagnetic transition.

\subsection{Elastic free energy}

Here we adopt a recently introduced model for square-to-rectangle martensite
  in two dimensions. While the elastic free energy has contributions arising 
from all the three  strain components,  namely the bulk dilational, 
deviatoric and shear strains, we has used only the deviatoric strain as the 
principal order parameter \cite{rajeev,kalaprl,kalaprb}. The effect of the
 other two strain components are included only in an indirect 
way (see below). The strain order parameter $\epsilon(\vec{r})$ is defined 
as \cite{rajeev,kalaprl,kalaprb}
\begin{equation}
\epsilon(\vec{r})=(\frac{\partial u_{x}(\vec{r})}{\partial x}-\frac{\partial
 u_y(\vec{r})}{\partial y})/\sqrt 2 = \epsilon_x(\vec{r}) - \epsilon_y(\vec{r}),
\end{equation}
where $u_x$ and $u_y$ are respectively the displacement fields in the $x$
 and $y$ directions.

The elastic free energy $F_e$ is the sum of  the local free-energy functional 
$F_L(\epsilon (\vec r))$ and  a nonlocal long-range term 
$F_{LR}(\epsilon (\vec r))$ that describes the transformation induced 
strain-strain interaction. The scaled form of the local free-energy $F_L$ 
is 
\begin{equation}
F_L=\int d\vec{r} \bigg[f_l(\epsilon(\vec{r})) + {D \over {2} }{({\nabla}
\epsilon(\vec{r}))^{2}} - \sigma (\vec {r})\epsilon(\vec {r}) \bigg].
\label{FL}
\end{equation}
where $D$ and $\sigma$ are in the scaled form.  In the above equation,
$f_l (\epsilon(\vec r))$ is the Landau polynomial for the first 
order transformation given by 
\begin{equation}
f_l(\epsilon(\vec {r}))={{\tau}\over{2}}{\epsilon(\vec r)}^2 - {\epsilon(\vec
 r)}^{4} + {{1}\over{2}}{\epsilon(\vec{r})}^{6}.
\end{equation}
Here, $\tau = (T - T_c)/(T_0 - T_c)$ is the scaled temperature. $T_0$ is the 
first-order transition temperature at which the free energy for the product
 and parent phases are equal, and $T_c$ is the  temperature below which 
there are only two degenerate global minima $\epsilon = \pm \epsilon_{M}$. 
 It is well known that nucleation occurs at localized defect sites in the 
crystal such as dislocations and grain boundaries \cite{ferra}. This  
 is modeled by  stress field \cite{Wang,rajeev,kalaprb} represented by the 
third term in Eq. \ref{FL}. This  modifies the free-energy $f_l$ in such a way 
that the austenite phase is locally unstable  leading to the nucleation of the 
product phase.

In principle, long-range interaction between the transformed domains results
 from the elimination of the shear and bulk strains. In our model, this term 
has been introduced in a phenomenological way even though  one can use the
 kernel obtained from St. Venant's principle \cite{kartha}.  This has been 
taken to have the form 
\begin{equation}
F_{LR}\{\epsilon\}=-{{1}\over{2}}\int \int d\vec{r}d\vec{r'}G(\vec{r}-\vec{r'})
\epsilon^{2}(\vec{r})
\epsilon^{2}(\vec{r'}).
\end{equation}
We note here that unlike the long range term emerging from the compatibility 
constraint which is linear in $\epsilon$, we have used $\epsilon^2$ term based 
on symmetry of the free energy under  $\epsilon \rightarrow -\epsilon$. 
However, we mention here that we have checked that using the kernel obtained  
by using compatibility condition  gives results similar to the kernel used 
here. The kernel $G(\vec{r}-\vec{r'})$ is best defined in the Fourier 
representation as
\begin{equation}
F_{LR}\{\epsilon\}={{1}\over{2}}\int d\vec{k} B\bigg({{\vec{k}}\over{k}}\bigg)
{\{ {\epsilon^{2}}(\vec{r})\}}_{k}
{\{ {\epsilon^{2}}(\vec{r})\}}_{k^{*}},
\end{equation}
where $\{ {\epsilon^{2}}(\vec{r})\}_{k}$ is the Fourier transform of
$\epsilon^{2}(\vec{r})$. 
The quantity ${\{ {\epsilon^{2}}(\vec{r})\}}_{k^{*}}$ is the
complex conjugate of  ${\{ {\epsilon^{2}}(\vec{r})\}}_{k}$.
The kernel $ B({{\vec{k}}/{k}} ) $, we use is designed to  pick up the correct 
habit plane directions which in the present case are $[11]$ and $[1\bar1]$, 
which are the favorable directions of growth of the product phase. In 
addition,   the free energy barrier is higher along the $[10]$ and $[01]$
 directions.  This is captured by the simple kernel given by
\begin{equation}
B\bigg({{\vec{k}}\over{k}}\bigg)= -\beta \theta(k-\Lambda){{\hat{k}}^{2}}_{x}
{{\hat{k}}^{2}}_{y},
\end{equation}
where $ {\hat{k}_{x}} $ and ${\hat{k}_{y}} $ are unit vectors in the
$x$ and $y$ directions. The step function $\theta(k-\Lambda)$
has been introduced to impose a cutoff on the range of the long-range
interactions. The constant $\beta$ is the strength of the interaction. 
The real space picture of $B({\vec{k}}/k)$ is similar to the long-range 
interaction of Kartha et. al \cite{kartha}.

We have used Rayleigh dissipation functional \cite{Landau} to account for the 
dissipation accompanying rapid interface motion described by 
\begin{equation}
R={{1}\over{2}}\gamma\int d\vec{r}{\big({{\partial }\over{\partial
t}}\epsilon (\vec{r},t)\big)}^{2}.
\end{equation}
This dissipation term has been successfully used to account for several 
features of  acoustic energy dissipated  during the martensitic transformation 
\cite{rajeev,kalaprl,kalaprb}. 

The Lagrangian then is given by $L = T - F$, where $F$ is the total  
free-energy  and $T $ is the kinetic energy associated with the system  is 
 given by 
\begin{equation}
T=\int d\vec{r}\rho\bigg[\bigg({{\partial u_{x}(\vec{r},t)}\over{\partial t}}
\bigg)^{2} + \bigg({{\partial u_{y}(\vec{r},t)}\over{\partial t}}\bigg)^{2}
\bigg].
\end{equation}
Here $\rho$ is the mass density. We use this for obtaining the equations of
 motion for $\epsilon$.

\subsection{Magnetic Free Energy}
As stated earlier, one of our main interest is to study the direction of 
magnetic order  in  the martensite variants and in the parent phase. However, 
it is well known that continuous spin ( $U(1)$ asymmetry) systems  do not 
support long range order in 2-D. For this reason, we use  a 4-state clock 
model. The free-energy has contributions  from exchange interaction, 
Zeeman energy, and  magneto-crystalline anisotropy energy and is  given by, 
\begin{eqnarray}
\nonumber f_m(\theta(\vec{r})) & = &-{j_p\over {2}} \sum_{\vec r'}  {\cos(\theta(\vec{r}) -  \theta(\vec{r'}))} \\ &- &{k_1 \over{2}} \cos (2\theta(\vec {r}) - \pi/4) 
  -  h{\cos(\theta(\vec {r}) - \phi) },
\end{eqnarray}
where the first term is the exchange energy, the second term is the magnetic 
anisotropic energy and the third the Zeeman term. The spin vector $\theta$ is 
allowed to  assume four values {\it i.e} $0$, $90$, $180$ and $270^o$ in the
austenite phase, and four similar values with respect to the diagonal  in 
the martensite phase (see below).  The sum for the first term is taken over 
the nearest neighbors. The coefficient $j_p$  is different for the austenite
and martensite phase ($p = a, m$).  $k_1$ is the anisotropy coefficient 
and $h$  the applied magnetic field.  ($\phi$ is the angle between the 
magnetic field and the $x-$ axis.) In the martensite phase,  we have used 
two different values for $j_m$, one each corresponding to the short  and 
long sides of the martensite variant ( in addition to the anisotropy term). 
The magnetic anisotropic energy sets the orientation of the magnetic order 
parameter along the easy axis. Usually, one uses $sin^2 \, \theta$, where $\theta$ is the angle measured with respect to the $x-$ axis.  From our earlier 
studies \cite{rajeev,kalaprl,kalaprb}, we know that the martensitic variants  
are along the diagonal.  Thus, in order to have  magnetic order parameter
oriented along these easy directions in the martensite phase, we have  
subtracted $45^o$. This gives upto an additive term $sin 2\,\theta$.
Lastly,  there is  magnetostriction energy  given by 
\begin{equation}
F_{me}(\epsilon(\vec{r}),\theta(\vec{r})) = -\int d\vec{r}{{b_1}\over{\sqrt 2}}
\epsilon(\vec{r})\sin(2\theta(\vec{r})),
\end{equation}
where $b_1$ is the phenomenological magnetoelastic constant.   The effect of 
the strain produced during the martensitic transformation on the magnetic 
transition is captured by this term. Here again, considering the fact that 
$\epsilon$ takes on positive and negative values for the two twin variants and 
that these twins are oriented along the diagonal, the above  term  is minimum 
along the diagonal.  The total magnetic contribution to the free energy is 
given by the sum of
 $F_m(\theta (\vec{r}))$ ($= \int f_m(\theta (\vec{r})) d \vec {r}$ ) and 
$F_{me}$.

The total free energy  is minimized with respect to both the order 
parameters $\epsilon$ and $\theta$. Here, we note that  while strain OP 
is continuous, the magnetic OP is discrete. Moreover, in our model, the 
martensitic transformation  is considered athermal and is therefore 
described by deterministic equations  obtained using 
\begin{equation}
{{d}\over{dt}}\bigg({{\delta L}\over{\delta {\dot{u}_i}}}\bigg)-
{{\delta L}\over{\delta u_{i}}}=-{ {\delta R}\over {\delta \dot{u}_i}},
 \, \, i = x, y.
\end{equation}
In terms of rescaled space and time variables, we get
\begin{eqnarray}
\nonumber {{{\partial}^{2} }\over{{\partial
t}^{2}}}\epsilon(\vec{r},t)& = & {\nabla}^2\bigg[{{\partial
f(\vec{r},t)}\over{\partial \epsilon(\vec{r},t)}}- \sigma(\vec{r})
- {\nabla}^2\epsilon(\vec{r},t)+
\gamma {{\partial }\over{\partial t}}\epsilon(\vec{r},t) \\
\nonumber & + &2\epsilon(\vec{r},t)\int d\vec{k}B(\vec
{k}/k)\{\epsilon^{2}(\vec{k},t)\}_{k}
 e^{i \vec {k} .\vec {r}} \\
& - & {b_1\over{\sqrt 2}} \sin(2\theta(\vec{r}) )  \bigg],
\label{epsilon}
\end{eqnarray}
Here, $\beta,\gamma$  and $b_1$ are  scaled parameters.   However, as the 
magnetic order parameter is allowed only four values ( Ising like), it is 
most conveniently dealt with using Metropolis Monte Carlo algorithm. 

\section{Numerical Simulations}
Numerical simulations to study the evolution of strain morphology and 
the magnetic ordering are  described below. We have solved 
 Eq. \ref{epsilon} using Euler's scheme with periodic boundary conditions 
after discretizing it on a $N \times N$ grid. The grid size  of the mesh used 
is $\Delta x= 1$  and the smallest time step is $\Delta t = 0.002$. 
Simulations  were carried out  for $N=128$ and $256$. However, the results  
reported here are for $N=128$. A pseudo-spectral technique is employed to 
compute the long-range term \cite{kalaprb}.  The cutoff $\Lambda$ in the 
long-range  expression is chosen to be $0.1$. The inhomogeneous stress field 
$\sigma(\vec{r})$ is appropriately chosen to describe the multi-defect 
configuration given by
\begin{equation}
\sigma(\vec{r})={\sum_j^{j_{max}}}\sigma_0(\vec{r_j}
)exp\bigg({{-|\vec{r}-\vec{r_j}|^{2}}\over{\zeta_j^{2}}}\bigg),
\label{defect}
\end{equation}
where $\sigma_0 (\vec {r}_j)$ is the magnitude of the stress field at sites 
$\vec{r}_j$ which are randomly chosen defect sites, $j_{max}$ is the total
 number of defect sites, and $\zeta_j$  the width of the field. A random 
distribution of $0.1\%$ defects is used which  for $N =128$ gives  
$j_{max}=16$.  The values of $\sigma_0(\vec{r_j})$ are taken to be uniformly 
distributed in  the interval $[-0.3,0.3]$. The parameters chosen for the 
martensite transformations are $\beta = 30$ and $\gamma = 1$. We start our 
simulations with the system being in a homogeneous state with 
$\epsilon(\vec{r},0)$ distributed in the interval $[-0.005,0.005]$.

The initial values of the spin  $\theta_{ij}$ at the location $ij$ are taken 
to be randomly  oriented among $0, 90, 180 $ or $270^o $ directions in the 
austenite phase and $45, 135, 225,$ or $315^o $ in the martensite phase.  
Thus, the allowed spin flips are through  $\pm 90^o$.  However, when crossing 
either the twin  or the martensite-austenite boundaries, we allow spin flips 
of  $\pm 45 ^o$ to ensure that  the spin orientations across the boundaries 
change appropriately.  The anisotropy coefficient is taken to be $k_1=0.6$ and 
the magnetostriction  coefficient $b_1=0.6$ in our  simulations 
(unless otherwise stated). The exchange interactions  $j_p$ which fixes the 
Curie temperatures of the magnetic phases   is chosen so as  to  obtain the 
required ordering of the Curie temperatures. In the Monte-Carlo simulations, 
when computing the equilibrium  magnetization at each temperature, the first 
$2.0 \times 10^5$
MC steps were discarded for thermalising, and averages are carried out over
 the next $1.0\times10^5$ MC steps.  ( We find that this duration is adequate 
for
 obtaining steady state values of the magnetization.) In the martensite phase, 
we have assumed that $j_m$
is different along the short and long axis of the variant denoted by 
 $j_{msh}$ and $j_{mlo}$ respectively. 

Before presenting our results, we note here that two types of simulations are 
possible due to separation of time scales associated with the two order 
parameters.  As mentioned earlier, our model describes  athermal martensite, 
and thus, the time evolution of the strain order parameter is considered fast
 compared to the magnetic order parameter (even though there is no direct 
mapping of Monte-Carlo time with  the real time).  Thus, the first possibility 
is to evolve  the strain order parameter  till a stationary morphology of the 
martensites is obtained and then a MC evolution of the magnetic order 
parameter is effected (or vice versa depending on the ordering of the 
transition and transformation temperatures). In the second type of simulation, 
the strain OP is allowed to develop for a short time (long enough time to 
reach stationary values) and MC evolution is carried out for about 
3,00,000 steps. This sequence is repeated till stationary values of both 
the OP's are  achieved. For most purposes, we find that the first type of 
simulations is adequate. However, we shall also report the results on the 
second type of simulation which demonstrates  the substantial mutual 
influence of the two order parameters.

\section{Results}
Before proceeding further, we note that it is necessary to  make a 
correspondence between the temperature scale used in the MC simulations with 
the temperature scale used in Eq. \ref{epsilon}. In our MC simulations, the 
scale of temperature is in the range of unity as we choose $j_a$ in the 
neighborhood of unity (in units of $k_B$) for the austenite phase.  However, 
the scaled temperature $\tau$, for the strain variable ranges  mostly over 
negative values. To map, we define a relation between the scaled temperature 
in MC simulations $\tau_{MC}$, and $\tau$ such that $\tau=0$ maps to 
$\tau_{MC} = 1$, ie.,  
$\tau_{MC} = \frac{\tau -(\tau_{Ms} - \tau_{Mf})}{\tau_{Ms} - \tau_{Mf}}$, 
where $\tau_{MC} = T/T_c$ with $T_c $ taken to be unity for a hypothetical 
austenite phase with $j_a = 0.885$. ( $\tau_{Ms}$ and $\tau_{Mf}$ are the 
scaled martensite start  and finish temperature, and the value of 
$\tau_{Ms} -\tau_{Mf} $ is determined numerically.) We have also used a 
small magnetic field along the positive  $x-$ direction to ensure that
 magnetization in the austenite phase is along the same  direction. We now 
consider a few cases of the ordering of 
temperatures listed earlier.

\subsection{Case: ${T_c}^A < T_{Ms} < {T_c}^M$}

\begin{figure}[h]
\includegraphics[height=5.5cm,width=8.5cm]{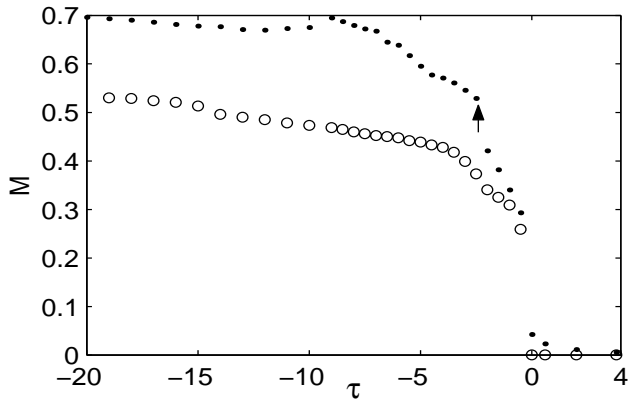}
\caption{Magnetization vs temperature plot for the condition ${T_c}^A < T_{ms}
 < {T_c}^M $.  }
\label{tm-amsm.ps}
\end{figure}

To obtain the above ordering of temperatures, we choose the exchange  
interactions in the austenite  phase to be  $j_a = 0.826$,  and for the 
martensite phase  $j_{msh} = 0.95$, $j_{mlo}= 0.35$ along the easy  and 
hard directions  respectively. With this, we get  ${T_c}^M = 1.0312$ 
( higher than $T_{Ms}$, ie., $\tau\sim 0.0$), which can only be determined  
numerically as it is influenced by $b_1$ and $k_1$. With the above value of 
$j_a$, we find ${T_c}^A = 0.937$ corresponding to $\tau = -1.98$ 
\cite{unreal}.   In this case, the Curie temperature of the martensite phase 
is not detectable as  $T_c^M$  is higher than $T_{Ms}$ and the high 
temperature austenite phase is paramagnetic. As the temperature is lowered 
below $T_{Ms}$ ( $\tau \sim 0.0$),  some regions  of the austenite phase 
 undergoes  structural transformation to the martensite phase which then 
orders ferromagnetically. This is reflected as an  abrupt  increase in the 
magnetization around $\tau = 0.0$  evident from Fig. \ref{tm-amsm.ps}.
 Further, magnetization grows in proportion to the  area fraction of the
 martensite phase until $T_c^A$ is reached.  At this point
 ($\tau  \approx -2.0$), one again finds  a sudden increase in the 
magnetization(Fig. \ref{tm-amsm.ps}, $\bullet$).  This can be attributed to
 ordering of the untransformed fraction of the austenite phase. In addition,
 we note that  the contribution to the total magnetization from the austenite
 phase is higher  as magnetization (per unit area) of this phase is oriented 
along the $x-$ direction compared to the martensite phase which is oriented 
along $\pm 45^o$.   However, as we decrease the temperature, the decreasing 
volume fraction of the austenite compensates the oriented  contribution from 
the austenite phase to the total magnetization. Consequently,  a further 
decrease in temperature leads to a marginal decrease in the magnetization
 around $\tau = -10.0$ eventually saturating to a value around 0.7. The
 sudden jump in the magnetization at 
$\tau = -2.0$ would not be seen if the  value of $T_c^A$ is such that the
 volume fraction of the untransformed austenite phase is small. In such a 
case, the magnetization curve increases smoothly  without such sharp changes. 
 This is shown on the same figure ( $\circ$ in Fig. \ref{tm-amsm.ps}) when
 $T_c^A = 0.765$ which corresponds to $\tau = -7.52$.  For this case,  the
 saturation value is much lower than for $\bullet$ as the parameter values 
are slightly different ($j_a=0.55,j_{msh}=0.90,j_{mlo}=0.35$ and $k_1=0.5$).
 On visual inspection, we find a large number of smaller sized twins which are 
not magnetized to the same extent as the twins for the case $\bullet$.

\begin{figure}[h]
\includegraphics[height=6.5cm]{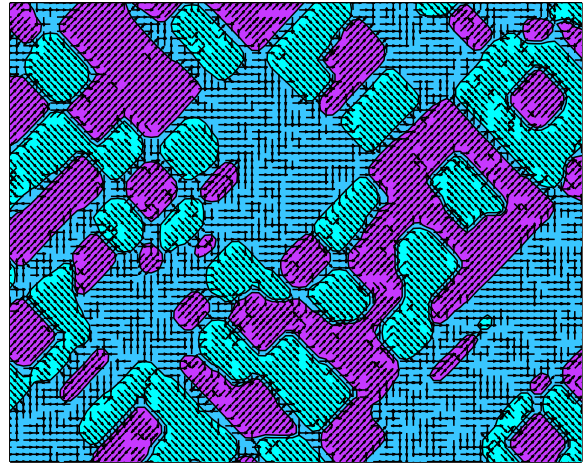}
\caption{(Online color) Morphology of the system at $\tau=-2.0$  for the case
${T_c}^M > \tau > {T_c}^A$. Here  pink and cyan corresponds to the two 
variants of the martensite and blue to the  austenite phase.  The magnetic
 ordering is present only in the martensite variants. For the sake of visual
 clarity a small region of  $85\times55$ grid points is displayed.}
\label{amsm-a} 
\end{figure}

\begin{figure}[h]
\includegraphics[height=6.5cm]{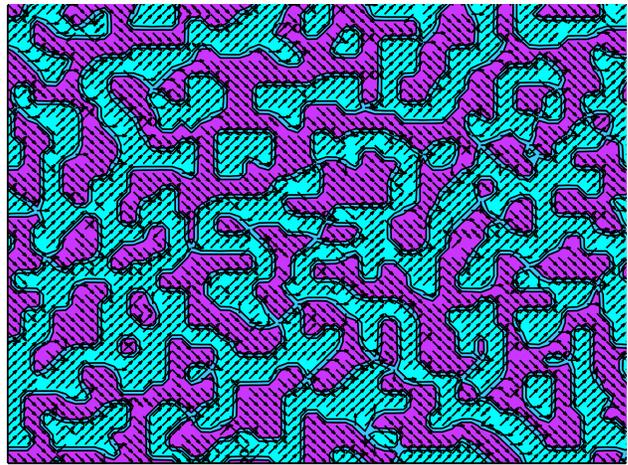}
\caption{(Online color) Morphology at $\tau = -9.0$ for the case 
${T_c}^M > \tau > {T_c}^A$ .  The system is fully 
transformed to martensite, and the spins are well ordered. 
A small region of $85\times55$ grid points is shown.}
\label{amsm-b}
\end{figure}

Figure \ref{amsm-a} shows  a morphological  plot of the magnetic ordering in 
the martensite domains   at $\tau = -2.0$ for the case shown in 
Fig. \ref{tm-amsm.ps} ($\bullet$).  Note that  the austenite phase is  
disordered. It is also clear from the Fig. \ref{amsm-a} that the two
 different variants  of the martensite phase have different orientations 
corresponding to the two easy directions of magnetization.  
 Figure \ref{amsm-b} shows the morphology for a lower temperature  
($\tau = -9.0$), a value at which the system has almost fully  transformed to 
the martensite phase  and the magnetization in both the twin  are well ordered
 even though the twin domains are rather small.

\subsection{Case: ${T_c}^M < T_{Ms} < {T_c}^A$}

As the temperature is lowered below $T_c^A$,  one first finds magnetic ordering
 of the austenite phase followed by the formation of the martensitic phase 
below $T_{Ms}$ and then  magnetic ordering of the martensite phase.  Choosing
 $j_a = 1.22$, one can  estimate  $T_c^A$ to be  1.38 which corresponds to 
$\tau_c^A = 12.28$. However, in our simulations, the transition temperature 
is identified by the peak in the susceptibility which in this case is seen at
 $\tau_c^A = 12.8$. (The shift in $\tau_c^A$ could arise from the finite sample
 size and small but finite magnetic field value of $h = 0.025$ used. It may
 be mentioned here that most measurements of magnetization in Heusler alloys
 are carried on with finite magnetic field. ) The martensite transition
 temperature obtained numerically is $T_c^M = 0.719$ for $j_{msh} =0.7$ and 
$j_{mlo} = 0.3$ which corresponds to  $\tau_{M}^c = -9.0$ \cite{unreal}.

\begin{figure}
\includegraphics[height=5.5cm,width=8.5cm]{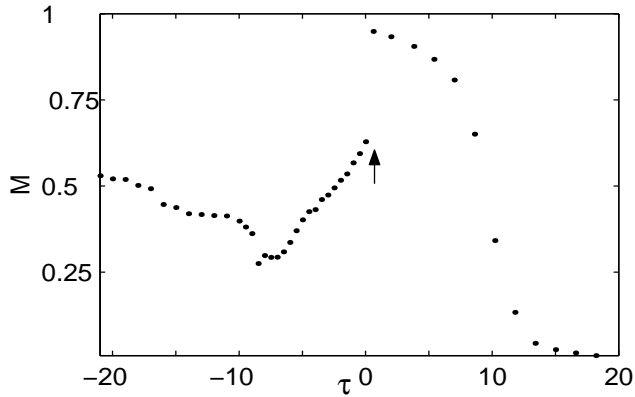}
\caption{Magnetization vs temperature plot for  $T_c^M < T_{Ms}< T_c^A$.}
\label{magausten}
\end{figure}

\begin{figure}[h]
\includegraphics[height=6.5cm]{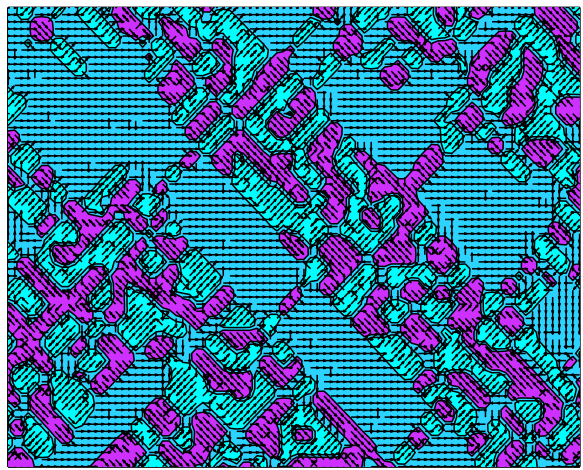}
\caption{(Online color) Ferromagnetic ordering of the austenite phase for the 
case $T_c^M < T_{Ms}< T_c^A$ at $\tau = -3$. Note that the spins in the 
austenite phase (blue color) are ordered and   the  martensite phase (pink and 
cyan corresponding to the two variants) is disordered. A small region of 
$85\times55$ grid points is shown.}
\label{austen-order}
\end{figure}

The magnetization curve for this case is  shown  in Fig. \ref{magausten}. In 
our simulations, the system shows a small amount magnetization  at 
$\tau = 12.8$ which  corresponds to the peak in the susceptibility. However,
 one observes a rapid increase in the magnetization $M$ as we cool further.  
From Fig. \ref{magausten}, it is clear that the austenite phase reaches near  
saturation value at $\tau = 0$ (  $\approx T_{Ms}$ ). As  the system is
 further cooled below $T_{Ms}$,  the martensite phase is formed  is
 not magnetically ordered at this temperature, which leads to a sudden drop in 
the magnetization. Indeed, the extent of the drop is equal to the area 
fraction of the austenite phase that transforms at $\tau =0$. On further 
cooling below 
$T_{c}^M$, as the martensite domains start magnetically ordered, the  
magnetization again increases reaching a value of $0.55$ as we approach 
$\tau = -20$. A snap shot of the spin ordering for $\tau = -3.0 $ is shown in
 Fig. \ref{austen-order}.  It is clear that  spins in the austenite phase are 
oriented along the $x-$ axis while those in the martensite phase are 
 disordered.  We note that the maximum value of the magnetization in  the 
transformed martensite phase (corresponding to $\tau = -20$) does not reach 
the saturation value $1/\sqrt 2$. This can be attributed to the substantial
 area fraction in between the martensite twins that is not ordered.  At these 
temperatures  even as the martensite domains quite fragmented, spins in the 
two martensite variants are ordered along their easy axis of magnetization, 
ie., $\pm 45^o$. The spin ordering here is similar to $T_c^M < T_{Ms}< T_c^A$ 
at similar temperature range Fig. \ref{amsm-b}.

\subsection{Case: ${T_c}^A < {T_c}^M < T_{Ms}$} 

\begin{figure}[b]
\includegraphics[height=5.5cm,width=8.5cm]{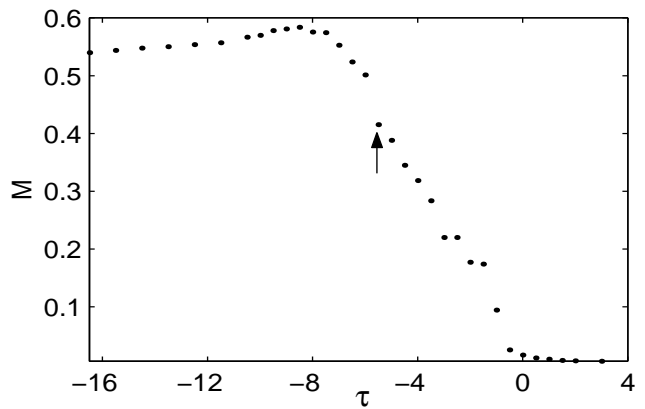}
\caption{Magnetization vs temperature plot for ${T_c}^A < {T_c}^M < T_{Ms}$.}
\label{t-mammsa}
\end{figure}

To obtain the required ordering of the transition and transformation 
temperatures, we choose the  exchange interaction  to be $j_a= 0.75$ and
 $j_{msh} = 0.9, j_{mlo} = 0.65$ for the austenite, and the  easy and hard
 directions of the martensite phase. With this choice, we get, 
 ${T_c}^A=0.83$ ($\tau = -5.5$ ) and ${T_c}^M = 0.98$ ( $\tau = -0.64$) for
 the austenite and the martensite phases respectively. Above  ${T_c}^M$, there 
is no ordering. As the temperature is  lowered below $T_c^M$, the magnetization
 increases till $T_c^A$ ( $\tau=-5.5$) at which one sees a small jump in the
 magnetization arising  from the fraction of the untransformed austenite phase
 as shown in Fig. 6. Thereafter, the magnetization increases to a maximum of
 0.6 and falls marginally. This decrease can be attributed to decreasing
 fraction of the austenite phase.

\subsection{Influence of magnetization on strain}

It is evident from Eq. 12,  that strain is affected by magnetization and vice 
versa when the system is in the martensite phase. This is a direct result of 
the magnetostriction coupling. We illustrate this interplay between the two 
order parameters for the case  $T_c^A < T_{Ms} < T_c^M$. As the high 
temperature phase is the paramagnetic austenite phase, we first obtain the 
initial stationary state for the strain order parameter by quenching the 
system below $T_{Ms}$ (ie.,  $\tau$ just less than 0.0 ).  In this  simulation,
 keeping  the initial state in the martensite phase, we allow the magnetic 
order to develop for a short time ( MC steps are typically 50,000 steps) 
followed by strain development.  This is repeated for several cycles. Both the 
strain and magnetic order parameters ( $M$ averaged over 30,000 MC steps)
 are monitored at the end of each cycle. We find that while maximum  changes 
occur in both the OP's  during the first few cycles, the magnetic order takes 
longer to reach a stationary value going through oscillatory changes.  The 
stationary values of the OP's are obtained after about 12 cycles. The 
percentage change at the end of the simulation in strain OP, with respect to 
the initial value is $\sim 6.0\%$. In contrast, the magnetic order changes by 
almost $\sim 10 \%$ with respect to the intial value, here taken as the value 
at the end of the first 50000 MC steps. 

\begin{figure}[t]
\includegraphics[height=5.0cm,width=4.2cm]{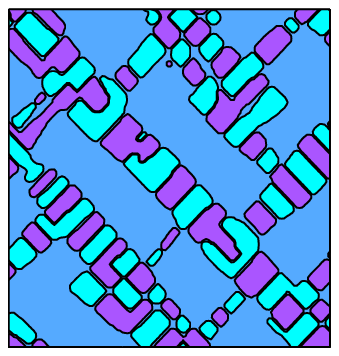}
\includegraphics[height=5.0cm,width=4.2cm]{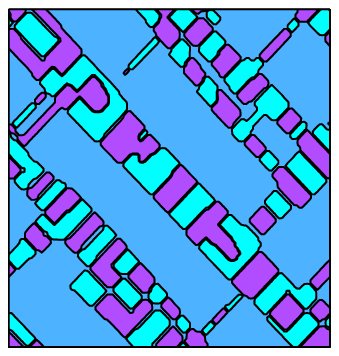}
\caption{(Online color) Strain morphology of the system  for 
${T_c}^A < T_{Ms} < {T_c}^M$ at $\tau=-0.5$ before the start (left) of the  
cycles and at the end configuration (right). The change in the area
 fraction of the  martensite phase due to the magnetization is clearly 
seen. The entire $128\times128$ grid points is shown.}
\label{diffmorph}
\end{figure}

\begin{figure}[h]
\includegraphics[height=5.5cm,width=8.5cm]{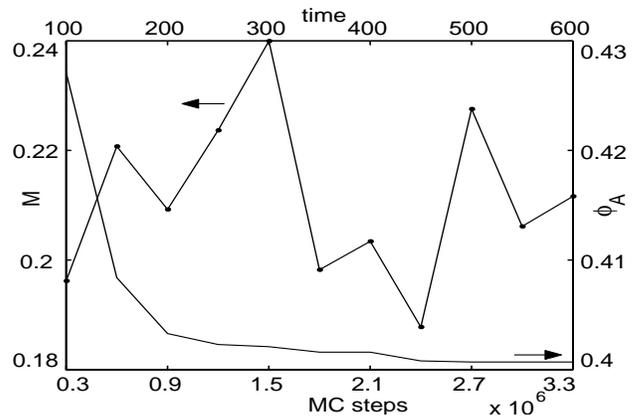}
\caption{Plot of magnetization and  area fraction versus MC steps and 
time for ${T_c}^A < T_{Ms} < {T_c}^M$ at $\tau=-0.5$.}
\label{fluct-mag}
\end{figure}

As the above simulations show that there is a strong mutual influence of the 
order parameters on each other, we have allowed the system to stabilize in each
 cycle., ie., strain is evolved till it reaches a stationary value  followed by
 the magnetic oder parameter reaching a stationary value consistent with the 
strain value. ( Near stationary $\phi_A$ values are reached in 25 time units.)
 Each magnetization run was for  3,00,000 MC steps and $M$ is calculated over 
the last 1,00,000 MC steps. This process is repeated for several cycles till
 stationary values of the OP's are obtained. Again, we found that while 
maximum  changes in the OP's occur in the first cycle, the magnetization 
which goes through oscillatory changes takes much longer time to stabilize. 
 The stationary values of the two order parameters obtained in this simulation
 are essentially the same as in the earlier simulation where the duration for
 the time development was much smaller. A plot of the strain and magnetization
 as a function of time and MC steps  is shown in Fig. \ref{fluct-mag}. Snap shots of the strain
 morphology at the end of first and last cycle are shown in Fig.  
\ref{diffmorph}. It is
 evident that the martensite boundaries have changed significantly from the 
initial configuration. However, we note that in real systems, during time 
evolution, there is a continuous feed back between the two order parameters.
 In this sense, simulations where the OP's are allowed to evolve for short 
durations are closer to continuous time development of both the OP's. 

\begin{figure}[h]
\includegraphics[height=5.5cm,width=8.5cm]{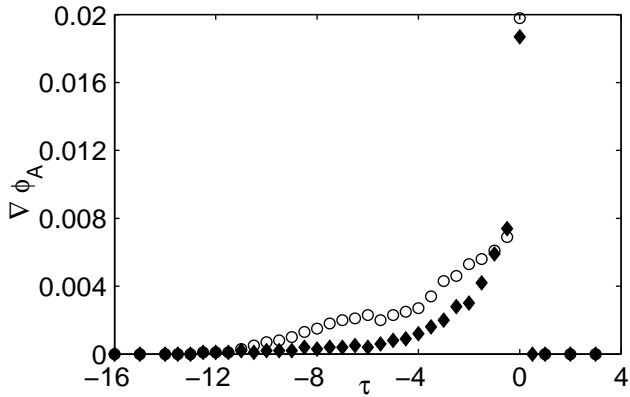}
\caption{Difference in area fraction vs scaled temperature, before and after 
magnetization for the cases $T_c^A < T_{Ms}<T_c^M$ ($\circ$) and 
$T_c^A<T_c^M<T_{Ms}$ ($\blacklozenge$).}
\label{straindiff}
\end{figure}

As the maximum change in strain occurs in the first cycle, we have examined the
 influence of temperature on this differential change. This is shown in Fig. 
\ref{straindiff}  for two conditions $T_c^A < T_{Ms}< T_c^M$ ( $\circ$) and 
$T_c^A < T_c^M< T_{Ms}$ ( $\blacklozenge$). As can be seen, in both cases, the 
differential change is maximum at the onset of the martensite phase.

\subsection{Hysteresis}

\begin{figure}[h]
\includegraphics[height=6.5cm]{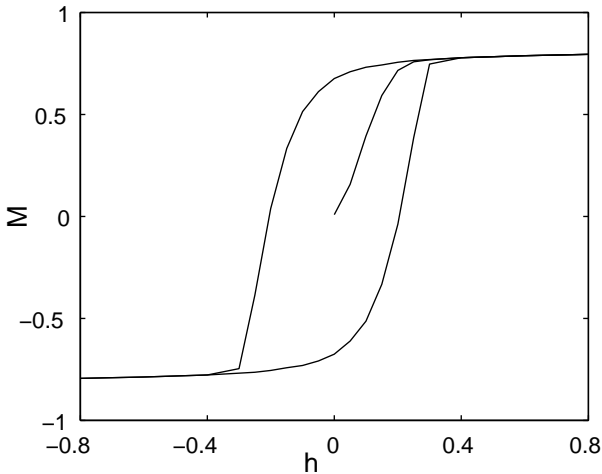}
\caption{Hysteresis plot for the case $T_c^A<T_{Ms}<T_c^M$ at $\tau=-3.0$.}
\label{hyst}
\end{figure}

We have also studied hysteresis loop keeping  the temperature  at an
 appropriate value. Here  the field is changed at a specified  rate from zero
 value to a positive maximum ( corresponding to the saturation value of the
 magnetization) and then reversing the field to the negative  maximum,  and 
back again to the positive maximum completing a cycle.  First, the system is 
thermalised to the desired temperature, here,  $\tau = -3.0$,  for the case
 ${T_c}^A < T_{Ms} < {T_c}^M$.  The loop is swept by varying the field at 
intervals of $h = 0.05$, except near saturation where it is varied at intervals
 of  $0.2$ ( beyond 0.6). Magnetization is evolved for  500 MC steps at each
 value of the field. At this temperature, $44\%$  of the system is transformed
 to  the martensite phase which is   ferromagnetic. Figure  \ref{hyst} shows
 the hysteresis cycle for this condition.

As is clear from the previous section, magnetization influences strain.
 However, as shown in our earlier work \cite{rajeev,kalaprl,kalaprb}, strain
 changes are always accompanied by  dissipation. For this reason, we have also
 monitored the dissipation accompanying strain during one cycle of  the
 external magnetic field. We find that the dissipation is  highest   when the 
field is close to  zero.   Here, it is pertinent to recall that in our
 earlier work we have shown that the dissipation accompanying the strain 
changes have all the characteristic features of acoustic emission
 \cite{rajeev,kalaprl,kalaprb}. If this interpretation is followed here, the 
dissipation is similar to the Barkhausen noise observed during the  hysteresis.

\begin{figure}[h]
\includegraphics[height=5.5cm,width=4.2cm]{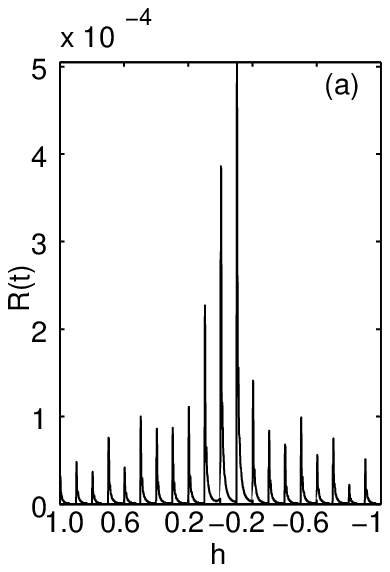}
\includegraphics[height=5.5cm,width=4.2cm]{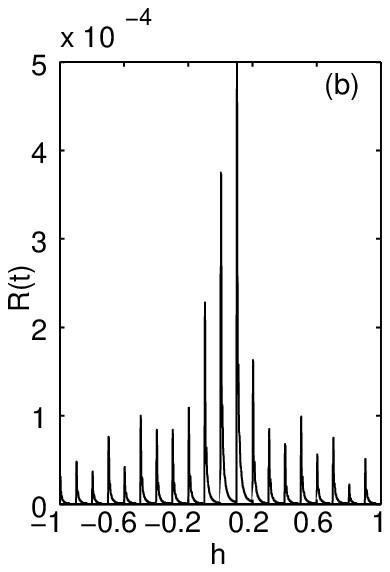}
\caption{Dissipation plotted as a function of external applied field for the 
case $T_c^A<T_{Ms}<T_c^M$ at $\tau=-3.0$. (a) Reverse cycle (from $h=1.0$ to 
$h=-1.0$ through zero filed), (b) Forward cycle.}
\label{dissp}
\end{figure}

\section{Summary and conclusions}

In summary, we have extended the recently introduced  two dimensional 
 square-to-rectangle martensite model to include the magnetic order parameter
 through a 4-state clock model to study  ferromagnetic martensites. The model 
includes  all the features of the structural transformation (ie., long-range 
 strain interaction, inertia, and defect centers that  act as nucleation sites)
 and magnetic transition (ie., the exchange interaction, anisotropy, and
 magnetostriction). The model is used to explain the magnetization curves of
 the austenite-martensite system under various ordering of the Curie
 temperatures and the martensite transformation temperature. By choosing 
appropriate  exchange interactions  in the  two  phases, we are able to obtain 
the characteristic magnetization plots. The model shows that the orientations 
of the magnetic order of the $\pm \epsilon$ twins are along $\pm 45^o$ 
directions, whereas it is  orientated along the $x -$ axis in the austenite 
phase. Further, the model  also captures  the abrupt change in the orientation 
of the magnetization across  the two variants of the martensite as also the 
austenite-martensite phase boundary.  The model shows that there is a subtle 
interplay between the growth of the strain  and magnetic order parameters that
 leads to  jumps in magnetization at $T_c^A$ in cases when the $T_c^A$ of the 
austenite phase is lower than $T_c^M$ ( or $T_{Ms}$), a feature  not predicted
 by the mean field models. The model also shows that the strain order 
parameter is influenced by the magnetic order and vice versa in a significant
 way. As 
this mutual influence is triggered by the magneto-striction coupling, the model
 may have relevance to Terfenol-D. In particular, the magnetic order appears to
 change the initial strain morpholgy of martensite domains in a substantial
 way.  Studies on  hysteresis coupled with the earlier identification of 
dissipation with acoustic emission shows that the dissipation is maximum
 while the applied field passes through zero value. This feature is similar
 to  the Barkhausen noise. The  Barkhausen noise has been explained as arising 
from  pinning and depinning of the magnetic domains. In our case, as shown 
earlier \cite{rajeev,kalaprl,kalaprb}, it results for the jerky motion of the
 interface. The driving force for this change is, however,  the interaction
 between the strain and the  magnetic order parameter. It is also clear that
 the approach taken here can be extended to systems where ferroelectric order
 parameter   has strong coupling to strain.


\begin{thebibliography}{00}
\bibitem{unreal} We stress that these unrealistic values of 
$j_{msh}$, $j_{mlo}$ and $j_a$ are necessary to have a good separation between 
$T_c^A$ and $T_c^M$. The anisotropy and the magnetoelastic  constant were 
taken to be $k_1 = 0.5$, and $b_1 = 0.6$ are also unrealistic.  These  were 
however, the minimum  values of the parameters for which the martensitic 
domains of reasonably small twins sizes  would  get ordered along the easy 
direction in a reasonable time.
\end{thebibliography}
\end{document}